\documentclass[aps,prb,twocolumn,groupedaddress,showpacs]{revtex4}
\usepackage{amsfonts}
\usepackage{amssymb}
\usepackage{amsmath}
\usepackage{color}
\usepackage{graphicx}

\begin{document}

\title{Effective speed of sound in phononic crystals}
\author{A.A. Kutsenko$^{a}$, A.L. Shuvalov$^{a}$, A.N. Norris$^{a,b}$}
\address{$^{a}\ $Universit\'{e} de Bordeaux, Institut de M\'{e}canique et
d'Ing\'{e}nierie de Bordeaux,
UMR 5295, Talence 33405, France, \
$^{b}$ Mechanical and Aerospace Engineering, Rutgers University,
Piscataway, NJ 08854, USA}
\begin{abstract}
A new formula for the effective quasistatic speed of sound $c$ in 2D
and 3D periodic materials is reported. The approach uses a
monodromy-matrix operator to enable direct integration in one of the
coordinates and exponentially fast convergence in others. As a
result, the solution for $c$ has a more closed form than previous
formulas. It significantly improves the efficiency and accuracy of
evaluating $c$ for high-contrast composites as demonstrated by a 2D
example with extreme behavior.
\end{abstract}

\maketitle

\section{Introduction}

Long-standing interest in modelling effective elastic properties of
composites with microstructure has  substantially intensified with the
emerging possibility of designing periodic structures in air \cite{DT} and
in solids \cite{MLWS} to form phononic crystals and other exotic
metamaterials, which open up exciting application prospects ranging from
negative index lenses to small scale multiband phononic devices\cite{V-H}.
This new prospective brings about the need for fast and accurate
computational schemes to test ideas \textit{in silico}. The most common
numerical tool is the Fourier or plane-wave expansion method (PWE). It is
widely used for calculating various spectral parameters including the
effective quasistatic speed of sound in acoustic \cite{KAG} and elastic \cite%
{QN} phononic crystals. At the same time, the PWE calculation is
known to face problems when applied to high-contrast composites
\cite{V-H}, which are of especial interest for applications.
Particularly riveting is the case where a soft ingredient is
embedded in a way breaking the connectivity of densely packed
regions of stiff ingredient. Physically speaking, the speed of
sound, which is large in a homogeneously stiff medium, should fall
dramatically when even a small amount of soft component forms a
'quasi-insulating network'. Note that this case, which implies a
strong effect of multiple interactions, is also ungainly for the
multiple-scattering approach \cite{DT,MLWS}.

The purpose of present Letter is to highlight a new method for evaluating
the quasistatic effective sound speed $c$ in 2D and 3D phononic crystals.
The idea is to recast the wave equation as a 1st-order 'ordinary'
differential system (ODS) with respect to one coordinate (say $x_{1}$) and
to use a monodromy-matrix operator defined as a multiplicative (or path)
integral in $x_{1}$.
By this means, we derive a formula for $c$ whose essential
advantages are an explicit integration in $x_{1}$ and an
exponentially small error of truncation in other coordinate(s). Both
these features of the analytical result are shown to significantly
improve the efficiency and accuracy of its numerical implementation
in comparison with the conventional PWE calculation, which is
demonstrated for a 2D steel/epoxy square lattice.   The power of the
new approach is especially apparent at high concentration $f$ of
steel inclusions, where the  effective speed $c$ displays a steep,
near vertical,  dependence for    $f\approx 1$,  a feature not
captured by conventional techniques like PWE.

\section{Effective speed: 2D acoustic waves\label{2D}}

\textbf{A. SETUP.} Consider the scalar wave equation
\begin{equation}
\pmb{\nabla }\cdot \left( \mu \pmb{\nabla }v\right) =-\rho \omega
^{2}v,  \label{0}
\end{equation}%
for time-harmonic shear displacement $v(\mathbf{x},t)=v(\mathbf{x}%
)e^{-i\omega t}$ in a 2D solid continuum\footnote{%
The subsequent results are equally valid for acoustic waves in fluid-like
phononic crystals under the standard interchange of $\rho $ and $\mu $ for
solids by $K^{-1}$ and $\rho ^{-1}$ for fluids.} with $\mathbf{T}$-periodic
density $\rho (\mathbf{x})$ and shear coefficient $\mu (\mathbf{x})$. Assume
a square unit cell $\mathbf{T}=\sum_{i}t_{i}\mathbf{a}_{i}=\left[ 0,1\right]
^{2}$ with unit translation vectors $\mathbf{a}_{1}\perp \mathbf{a}_{2}$
taken as the basis for $\mathbf{x}=\sum_{i}x_{i}\mathbf{a}_{i}.$ Imposing
the Floquet condition $v(\mathbf{x})=u(\mathbf{x})e^{i\mathbf{k\cdot x}}$
where $u(\mathbf{x})$ is periodic and $\mathbf{k}=k\pmb{\kappa }$ ($%
\left\vert \pmb{\kappa }\right\vert =1$), Eq.\ (\ref{0}) becomes%
\begin{gather}
({\mathcal{C}}_{0}+{\mathcal{C}}_{1}+{\mathcal{C}}_{2})u=\rho \omega
^{2}u\ \ \mathrm{with\ }{\mathcal{C}}_{0}u=-\pmb{\nabla }(\mu
\pmb{\nabla }u),
\notag \\
{\mathcal{C}}_{1}u=-i\mathbf{k}\cdot (\mu \pmb{\nabla }u+\pmb{\nabla }%
(\mu u)),\ \ {\mathcal{C}}_{2}u=k^{2}\mu u.  \label{501}
\end{gather}%
Regular perturbation theory applied to (\ref{501}) yields the effective
speed $c(\pmb{\kappa })=\lim_{\omega ,k\rightarrow 0}\omega (\mathbf{k})/k$
in the form \cite{KSAP}%
\begin{align}\label{6}
&c^{2}(\pmb{\kappa })=\mu _{\mathrm{eff}}(\pmb{\kappa })/\langle \rho \rangle
,\ \mu _{\mathrm{eff}}(\pmb{\kappa })=\left\langle \mu \right\rangle -M(%
\pmb{\kappa })\ \mathrm{with}   \\
&M(\pmb{\kappa })=%
\begin{matrix}
\sum\nolimits_{i,j=1}^{2}%
\end{matrix}%
M_{ij}\kappa _{i}\kappa _{j},\ M_{ij}=\left( {\mathcal{C}}_{0}^{-1}\partial
_{i}\mu ,\partial _{j}\mu \right) =M_{ji}, \notag
\end{align}%
where $\partial _{i}\equiv \partial /\partial x_{i}$, spatial
averages are defined by
\begin{equation}
\left\langle f\right\rangle \equiv
\begin{matrix}
\int_{\mathbf{T}}%
\end{matrix}%
f(\mathbf{x})\mathrm{d}\mathbf{x\ }\ \big(=\left\langle \left\langle
f\right\rangle _{1}\right\rangle _{2},\mathbf{\ }\left\langle f\right\rangle
_{i}\equiv
\begin{matrix}
\int_{0}^{1}%
\end{matrix}%
f(\mathbf{x})\mathrm{d}x_{i}\big),  \label{6.1}
\end{equation}%
and $(\cdot ,\cdot )$ denotes the scalar product in
$L^{2}(\mathbf{T})$ so that $(f,h)=\left\langle fh^{\ast
}\right\rangle $ ($^{\ast }$ means complex conjugation). The
difficulty with (\ref{6}) is that it involves the inverse of a
partial differential operator ${\mathcal{C}}_{0}$. One solution is
to apply a double Fourier expansion to ${\mathcal{C}}_{0}^{-1}$ and
$\partial
_{i}\mu $ in (\ref{6}). This leads to the PWE formula for the effective speed%
\cite{KAG} which is expressed via infinite vectors and the inverse of the
infinite matrix of Fourier coefficients of $\mu (\mathbf{x})$. Numerical
implementation of the PWE formula requires dealing with large dense
matrices, especially in the case of high-contrast composites for which the
PWE convergence is slow (see \S \ref{N}). An alternative "brute force"
procedure of the scaling approach is to numerically solve the partial
differential equation ${\mathcal{C}}_{0}h=\partial _{i}\mu $ for the
$\mathbf{1}$-periodic function $h(\mathbf{x})$ (e.g. via the boundary integral
method\cite{AADW}).

The new approach proposed here leads to a more efficient formula for $c$
based on direct analytical integration in one coordinate direction. There
are two ways of doing so. The first proceeds from the ODS form of the wave
equation (\ref{0}) itself, which means 'skipping' (\ref{6}). This is
convenient for deriving $c(\pmb{\kappa })$ in the principal directions $%
\pmb{\kappa }\parallel \mathbf{a}_{1,2},$ see \S IIB. The second method is
more closely related to the conventional PWE and scaling approaches in that
it also starts from (\ref{6}) but treats it differently, namely,
the equation ${\mathcal{C}}_{0}h=\partial _{i}\mu $ is cast in
ODS form and analytically integrated in one coordinate.
This is basically equivalent to the former method, but enables an easier
derivation of the off-diagonal component $M_{12}$ for the anisotropic case,
see \S IIC.

\medskip
\textbf{B. Wave speed in the principal directions. }The wave equation (\ref%
{0}) may be recast as
\begin{gather}
\pmb{\eta }^{\prime }=\mathcal{Q}\pmb{\eta }\quad \mathrm{with}\ \ \mathcal{A=}-\partial _{2}(\mu \partial _{2})
,~  \notag \\
\mathcal{Q}=%
\begin{pmatrix}
0 & \mu ^{-1} \\
\mathcal{A}-\rho \omega ^{2} & 0%
\end{pmatrix}%
,\
\pmb{\eta }%
(\mathbf{x})=%
\begin{pmatrix}
v \\
\mu v^{\prime }%
\end{pmatrix}%
,\   \label{M3}
\end{gather}%
where 
$\prime $ stands for $\partial _{1}$. The solution to
Eq.\ (\ref{M3}) for initial data $\pmb{\eta }(0,x_{2})\equiv \pmb{\eta }%
(0,\cdot )$ at $x_{1}=0$ is
\begin{gather}
\pmb{\eta }(x_{1},\cdot )=\mathcal{M}\left[ x_{1},0\right] \pmb{\eta }%
(0,\cdot )\ \mathrm{with}  \notag \\
\mathcal{M}\left[ a,b\right] =%
\begin{matrix}
\widehat{\int }_{b}^{a}%
\end{matrix}%
(\mathcal{I}+\mathcal{Q}\mathrm{d}x_{1}).  \label{M5}
\end{gather}%
The operator $\mathcal{M}\left[ x_{1},0\right] $ is formally the matricant,
or propagator, of (\ref{M3}) defined through the multiplicative integral $%
\widehat{\int }$ (with $\mathcal{I}$ denoting the identity operator).
It is assumed for the moment that $\rho (\mathbf{x})$ and $\mu (\mathbf{x})$
are smooth to ensure the existence of $\mathcal{M}$. The matricant over a
period, $\mathcal{M}\left[ 1,0\right] $, is called the monodromy matrix.

Assume the Floquet condition with the wave vector $\mathbf{k}=(k_{1}\ 0)^{%
\mathrm{T}}$ so that $v(\mathbf{x})=u(\mathbf{x})e^{ik_{1}x_{1}}$ and  $
\pmb{\eta }(1,\cdot )$ $=\pmb{\eta }(0,\cdot )e^{ik_{1}}.$ By (\ref{M5})$_{1}
$, this implies the eigenproblem
\begin{equation}
\mathcal{M}\left[ 1,0\right] \mathbf{w}(k_{1})=e^{ik_{1}}\mathbf{w}(k_{1}).
\label{2}
\end{equation}%
where $\mathcal{M}\left[ 1,0\right] $ depends on $\omega $. Eq.\ (\ref{2})
defines $k_{1}=k_{1}(\omega )$ and hence $\omega =\omega (k_{1}),$ where $%
\omega ^{2}$ is the eigenvalue of (\ref{0}) with $v(\mathbf{x})=u(\mathbf{x}%
)e^{ik_{1}x_{1}}$. The effective speed $c(\kappa _{1})=\lim_{\omega
,k_{1}\rightarrow 0}\omega /k_{1}$ can therefore be determined by applying
perturbation theory  to (\ref{2}) as $\omega ,k_{1}\rightarrow 0$. The
asymptotic form of $\mathcal{M}\left[ 1,0\right] $  follows from
definitions (\ref{M3}) and (\ref{M5})$_{2}$ as
\begin{gather}
\mathcal{M}\left[ 1,0\right] =\mathcal{M}_{0}+\omega ^{2}\mathcal{M}%
_{1}+O(\omega ^{4})\ \mathrm{where:}  \notag \\
\mathcal{M}_{0}\equiv \mathcal{M}_{0}\left[ 1,0\right] ,\ \mathcal{M}_{0}%
\left[ a,b\right] =%
\begin{matrix}
\widehat{\int }_{b}^{a}%
\end{matrix}%
(\mathcal{I}+\mathcal{Q}_{0}\mathrm{d}x_{1})\ \mathrm{with}\   \notag \\
\mathcal{Q}_{0}\equiv \mathcal{Q}_{\omega =0}=%
\begin{pmatrix}
0 & \mu ^{-1} \\
\mathcal{A} & 0%
\end{pmatrix}%
,  \label{M6} \\
\mathcal{M}_{1}=\int_{0}^{1}\mathcal{M}_{0}\left[ 1,x_{1}\right]
\begin{pmatrix}
0 & 0 \\
-\rho  & 0%
\end{pmatrix}%
\mathcal{M}_{0}\left[ x_{1},0\right] \mathrm{d}x_{1}.  \notag
\end{gather}%
Note the identities $\mathcal{Q}_{0}\mathbf{w}_{0}=\mathbf{0},\ \mathcal{Q}%
_{0}^{+}\widetilde{\mathbf{w}}_{0}=\mathbf{0}$ and hence%
\begin{gather}
\mathcal{M}_{0}\left[ a,b\right] \mathbf{w}_{0}=\mathbf{w}_{0},\ \mathcal{M}%
_{0}^{+}\left[ a,b\right] \widetilde{\mathbf{w}}_{0}=\widetilde{\mathbf{w}}%
_{0}\mathbf{\ }\ (\forall a,b)  \notag \\
\mathrm{for}\ \mathbf{w}_{0}=(1\ 0)^{\mathrm{T}},\ \widetilde{\mathbf{w}}%
_{0}=(0\ 1)^{\mathrm{T}}.  \label{1}
\end{gather}%
By (\ref{1})$_1$ $\mathbf{w}_{0}$ is an eigenvector of
$\mathcal{M}_{0}$ with the eigenvalue 1, and it can   be shown to be
a single eigenvector.
Therefore $\mathbf{w}(k_{1})=\mathbf{w}_{0}+k_{1}\mathbf{w}_{1}+k_{1}^{2}%
\mathbf{w}_{2}+O(k_{1}^{3})$ and$\ \omega =ck_{1}+O(k_{1}^{2})$. Insert
these expansions along with (\ref{M6})$_{1}$ in (\ref{2}) and collect the
first-order terms in $k_{1}$ to obtain
\begin{equation}
\mathcal{M}_{0}\mathbf{w}_{1}=\mathbf{w}_{1}+i\mathbf{w}_{0}\ \Rightarrow \
\mathbf{w}_{1}=i(\mathcal{M}_{0}-\mathcal{I})^{-1}\mathbf{w}_{0}.  \label{3}
\end{equation}%
According to (\ref{1}), $\mathcal{M}_{0}-\mathcal{I}$ has no inverse but is
a one-to-one mapping from the subspace orthogonal to $\mathbf{w}_{0}$ onto
the subspace orthogonal $\widetilde{\mathbf{w}}_{0};$ hence, $\mathbf{w}_{1}$
exists and $\widetilde{\mathbf{w}}_{0}\cdot \mathbf{w}_{1}$ is uniquely
defined. The terms of second-order in $k_{1}$ in (\ref{2}) then imply
\begin{equation}
\mathcal{M}_{0}\mathbf{w}_{2}+c^{2}\mathcal{M}_{1}\mathbf{w}_{0}=i\mathbf{w}%
_{1}+\mathbf{w}_{2}.  \label{4}
\end{equation}%
Scalar multiplication on both sides by $\widetilde{\mathbf{w}}_{0}$ leads,
with account for (\ref{1}) and (\ref{M6})$_{4}$, to $c^{2}\langle \rho
\rangle =-i\left\langle \widetilde{\mathbf{w}}_{0}\cdot \mathbf{w}%
_{1}\right\rangle _{2}$, whence by (\ref{3}$_{2}$)
\begin{equation}
c^{2}(\kappa _{1})=\ \langle \rho \rangle ^{-1}\left\langle \widetilde{%
\mathbf{w}}_{0}\cdot (\mathcal{M}_{0}-\mathcal{I})^{-1}\mathbf{w}%
_{0}\right\rangle _{2},  \label{5}
\end{equation}%
where the notation $\left\langle \cdot \right\rangle _{2}$ is explained in (%
\ref{6.1}). Interchanging variables $x_{1}\rightleftarrows x_{2}$ in the
above derivation yields a similar result for $c(\kappa _{2})$ as follows
\begin{gather}
c^{2}(\kappa _{2})=\left\langle \rho \right\rangle ^{-1}\left\langle
\widetilde{\mathbf{w}}_{0}\cdot (\widetilde{\mathcal{M}}_{0}-\mathcal{I}%
)^{-1}\mathbf{w}_{0}\right\rangle _{1}\ \ \mathrm{where}  \notag \\
\widetilde{\mathcal{M}}_{0}=%
\begin{matrix}
\widehat{\int }_{0}^{1}%
\end{matrix}%
(\mathcal{I}+\widetilde{\mathcal{Q}}_{0}\mathrm{d}x_{2}),  \label{f6} \\
\widetilde{\mathcal{Q}}_{0}=%
\begin{pmatrix}
0 & \mu ^{-1} \\
\widetilde{\mathcal{A}} & 0%
\end{pmatrix}%
,\ \ \widetilde{\mathcal{A}}=-\partial _{1}\mu \partial _{1}.  \notag
\end{gather}%
The result for a rectangular lattice with $\mathbf{T}=\left[ 0,T_{1}\right]
\times \left[ 0,T_{2}\right] $ is obtained by replacing $x_{i}$ with $%
x_{i}/T_{i}.$

\medskip
\textbf{C. The full matrix} $M_{ij}$. The anisotropy of the effective speed $%
c(\pmb{\kappa }),$ i.e. its dependence on the wave normal $\pmb{\kappa }%
\equiv \mathbf{k/}k$, is determined by the quadratic form $M(\pmb{\kappa }%
)=\sum\nolimits_{i,j=1}^{2}M_{ij}\kappa _{i}\kappa _{j}$ (see Eq.\ (\ref{6}%
)), and represented by the ellipse of (squared) slowness $c^{-2}(\mathbf{%
\kappa })$. Eqs.\ (\ref{5}) and (\ref{f6})$_{1}$, which define $c(\kappa
_{i})$ and so $M_{ii}$, suffice for the case where $\mathbf{T}$ is
rectangular and $\mu (\mathbf{x})$ is even in (at least) one of $x_{i}$ so
that the effective-slowness ellipse is $c^{-2}(\pmb{\kappa }%
)=\sum\nolimits_{i=1,2}c^{-2}(\kappa _{i})\kappa _{i}^{2}$ with the
principal axes parallel to $\mathbf{a}_{1}\perp \mathbf{a}_{2}$. Otherwise $%
c(\pmb{\kappa })$ for arbitrary $\pmb{\kappa }$ requires finding the
off-diagonal component $M_{12}$. For this purpose, with reference to (\ref{6}%
), consider the equation
\begin{equation}
{\mathcal{C}}_{0}h=\partial _{1}\mu   \label{7.0}
\end{equation}%
for $\mathbf{1}$-periodic $h(\mathbf{x})$. With the above notations this can
be written as $-(\mu h^{\prime })^{\prime }+\mathcal{A}h=\mu ^{\prime }$ or,
more conveniently, $(\mu \widetilde{h}^{\prime })^{\prime }=\mathcal{A}%
\widetilde{h}$\ with \ $\widetilde{h}=h+x_{1}$. The latter is equivalent to
\begin{equation}
\pmb{\xi }^{\prime }=\mathcal{Q}_{0}\pmb{\xi }\mathrm{\ \ where\
}\pmb{\
\xi }=%
\begin{pmatrix}
h+x_{1} \\
\mu (h^{\prime }+1)%
\end{pmatrix}
\label{7}
\end{equation}%
and $\mathcal{Q}_{0}$ is given in (\ref{M6})$_{3}$. The general solution to (%
\ref{7}) is
\begin{equation}
\pmb{\xi }(x_{1},\cdot )=\mathcal{M}_{0}\left[ x_{1},0\right] \pmb{\xi }%
(0,\cdot ),  \label{8}
\end{equation}%
where $\mathcal{M}_{0}\left[ x_{1},0\right] $ is defined in (\ref{M6})$_{2}$%
, and $\pmb{\xi }(0,\cdot )$ is the initial data at $x_{1}=0$. The
periodicity of $h$ implies $\pmb{\xi }(1,\cdot )=\pmb{\xi }(0,\cdot )+%
\mathbf{w}_{0},$ while $\pmb{\xi }(1,\cdot )=\mathcal{M}_{0}\pmb{\xi }%
(0,\cdot )$ by (\ref{8}). Hence $\pmb{\xi }(0,\cdot )=(\mathcal{M}_{0}-%
\mathcal{I})^{-1}\mathbf{w}_{0}$ and so (\ref{7.0}) is solved by
\begin{equation}
\pmb{\xi }(x_{1},\cdot )=\mathcal{M}_{0}\left[ x_{1},0\right] (\mathcal{M}%
_{0}-\mathcal{I})^{-1}\mathbf{w}_{0}.  \label{9}
\end{equation}%
Substituting (\ref{9}) into the definition of $M_{12}$ in (\ref{6}) yields
\begin{align}
M_{12}& =(\mathcal{C}_{0}^{-1}\partial _{1}\mu ,\partial _{2}\mu )=\langle
h\partial _{2}\mu \rangle =\left\langle \partial _{2}\mu \mathbf{w}_{0}\cdot %
\pmb{\xi }\right\rangle   \notag \\
& =\left\langle \partial _{2}\mu \mathbf{w}_{0}\cdot \mathcal{M}%
_{0}[x_{1},0](\mathcal{M}_{0}-\mathcal{I})^{-1}\mathbf{w}_{0}\right\rangle .
\label{f5}
\end{align}

Note that the formula (\ref{f5}) for $M_{12}$ requires more computation than
the formulas (\ref{5}) and (\ref{f6})$_{1}$ for $M_{ii}$. Interestingly, if
the unit cell $\mathbf{T}$ is square, then, for an arbitrary (periodic) $\mu
(\mathbf{x})$, Eq.\ (\ref{f5}) can be circumvented by using the identity $%
M_{12}=(\widetilde{M}_{11}-\widetilde{M}_{22})/2$, where $\widetilde{M}_{ii}$
follow from Eqs.\ (\ref{5}) and (\ref{f6})$_{1}$ applied to the square
lattice obtained from the given one by turning it 45$^{\circ }$.

\medskip
\textbf{D. Discussion. }The two lines of attack outlinedmentioned in
\S II.A are equivalent in that the formula (\ref{5}) for the
effective speed $c(\kappa _{1})$ in the principal direction can also
be inferred from Eq.\ (\ref{6}).
Inserting the solution (\ref{9}) of (\ref{7.0}) defines the component $%
M_{11} $ as
\begin{equation}
M_{11}=\big({\mathcal{C}}_{0}^{-1}\partial _{1}\mu ,\partial _{1}\mu \big)%
=\langle h\mu ^{\prime }\rangle =\langle \mu ^{\prime }{\mathbf{w}}_{0}\cdot %
\pmb{\xi }\rangle -\langle x_{1}\mu ^{\prime }\rangle .  \label{f1}
\end{equation}%
Integrating by parts each term in the last identity and using the
periodicity of $\mu (\mathbf{x})$ along with Eqs.\ (\ref{M6})$_{3}$, (\ref{1}%
), (\ref{7})-(\ref{9}) (see also the notation (\ref{6.1})) yields
\begin{align}
& \langle \mu ^{\prime }{\mathbf{w}}_{0}\cdot \pmb{\xi }\rangle =-\langle
\widetilde{\mathbf{w}}_{0}\cdot (\mathcal{M}_{0}-\mathcal{I})^{-1}\mathbf{w}%
_{0}\rangle _{2}+\langle \mu (0,x_{2})\rangle _{2},  \notag \\
& -\langle x_{1}\mu ^{\prime }\rangle =\langle \mu \rangle -\langle \mu
(1,x_{2})\rangle _{2}=\langle \mu \rangle -\langle \mu (0,x_{2})\rangle _{2}.
\label{f3}
\end{align}%
%
%
%
%
%
%
%
%
Thus, $M_{11}$$=\langle \mu \rangle $$-\left\langle \widetilde{\mathbf{w}}%
_{0}\cdot (\mathcal{M}_{0}-\mathcal{I})^{-1}\mathbf{w}_{0}\right\rangle _{2}$
which leads to (\ref{5}), QED. Note  that Eq.\ (\ref{f5}) is also obtainable
via the monodromy matrix of the wave equation (\ref{0}) (the approach of \S %
IIB) with $v(\mathbf{x})=u(\mathbf{x})e^{i\mathbf{k\cdot x}}$ and $\mathbf{%
k\nparallel a}_{i},$ but this method of derivation of $M_{12}$ is
lengthier than in \S IIC.


As another remark, it is instructive to recover a known result for the case
where $\mu (\mathbf{x})$ is periodic in one coordinate and does not depend
on the other, say $\mu (x_{1},x_{2})=\mu (x_{1})$. Using (\ref{M6})$_{2}$, (%
\ref{M6})$_{3}$ and (\ref{f6})$_{3}$ gives
\begin{equation}
(\mathcal{M}_{0}-\mathcal{I})%
\begin{pmatrix}
0 \\
\langle \mu ^{-1}\rangle _{1}^{-1}%
\end{pmatrix}%
=\mathbf{w}_{0},~(\widetilde{\mathcal{M}}_{0}-\mathcal{I})%
\begin{pmatrix}
0 \\
\mu (x_{1})%
\end{pmatrix}%
=\mathbf{w}_{0}.  \label{f9}
\end{equation}%
Therefore, by (\ref{5}) and (\ref{f6})$_{1}$, $c^{2}(\kappa _{1})=\langle
\mu ^{-1}\rangle _{1}^{-1}/\left\langle \rho \right\rangle $ and $%
c^{2}(\kappa _{2})=\langle \mu \rangle _{1}/\left\langle \rho \right\rangle $
while $M_{12}=0$ by (\ref{f5}) with $\partial _{2}\mu =0$.

Finally, we note that, while the above evaluation of quasistatic speed $c$
is exact, using the same monodromy-matrix approach also provides a
closed-form approximation of $c.$ For the isotropic case, it is as follows
(see \cite{KSAP} for more details):%
\begin{equation}
c^{2}\approx \frac{1}{2\left\langle \rho \right\rangle }\left( \left\langle
\left\langle \mu ^{-1}\right\rangle _{1}^{-1}\right\rangle _{2}+\left\langle
\left\langle \mu \right\rangle _{2}^{-1}\right\rangle _{1}^{-1}\right) .
\label{M18}
\end{equation}

\section{Effective speeds in principal directions for 3D elastic waves}

The equation for time-harmonic elastic wave motion $\mathbf{v}(\mathbf{x},t)$%
$=$$\mathbf{v}(\mathbf{x})e^{-i\omega t}$ is, with repeated suffices summed,
\begin{equation}
-\partial _{j}(c_{ijkl}\partial _{l}v_{k})=\rho \omega ^{2}v_{i}\
(i,j,k,l=1,2,3),  \label{3D1}
\end{equation}%
where density $\rho (\mathbf{x})$ and compliances $c_{ijkl}(\mathbf{x})$ are
$\mathbf{T}$-periodic in a 3D periodic medium. Assume a cubic unit cell $%
\mathbf{T}=\sum_{i}t_{i}\mathbf{a}_{i}=\left[ 0,1\right] ^{3}$ and refer the
components $x_{i},~v_{i}$ and $c_{ijkl}$ to the orthogonal basis formed by
the translation vectors $\mathbf{a}_{i}$. Impose the condition $\mathbf{v}(%
\mathbf{x})=\mathbf{u}(\mathbf{x})e^{i\mathbf{k\cdot x}}$ with periodic $%
\mathbf{u}(\mathbf{x})=(u_{i})$ and take $\mathbf{k}$ parallel to one of $%
\mathbf{a}_{i}$, e.g. to $\mathbf{a}_{1}$. Eq.\ (\ref{3D1}) may be rewritten
in the form
\begin{gather}
\pmb{\eta }^{\prime }=\mathcal{Q}\pmb{\eta }\ \ \mathrm{with}\ \ \pmb{\eta }(%
\mathbf{x})=%
\begin{pmatrix}
(u_{i}) \\
(c_{i1kl}\partial _{l}u_{k})%
\end{pmatrix}%
,  \notag \\
\mathcal{Q}=%
\begin{pmatrix}
-\mathcal{C}^{-1}\mathcal{A}_{1} & \mathcal{C}^{-1} \\
\omega ^{2}\rho \delta _{ij}+\mathcal{A}_{2}-\mathcal{A}_{1}\mathcal{C}^{-1}%
\mathcal{A}_{1} & \mathcal{A}_{1}\mathcal{C}^{-1}%
\end{pmatrix}
\label{3D2}
\end{gather}%
where the self-adjoint matrix operators $\mathcal{C}$ and $\mathcal{A}_{1,2}$
are 
\begin{gather}
\mathcal{C}=(c_{i1k1}),\ \ \mathcal{A}_{1}(u_{i})=(c_{i1ka}\partial
_{a}u_{k}),  \notag \\
\mathcal{A}_{2}(u_{i})=(\partial _{a}(c_{iakb}\partial _{b}u_{k}))\ \mathrm{%
with}\ \ \ a,b=2,3.  \label{3D3}
\end{gather}%
Like in the 2D case, denote the monodromy matrix for (\ref{3D2}) at $\omega
=0$ by $\mathcal{M}_{0}=\widehat{\int }_{0}^{1}(\mathcal{I}+\mathcal{Q}%
_{0}dx_{1})$ where $\mathcal{Q}_{0}=\mathcal{Q}_{\omega =0}$, and also
introduce the 6$\times $3 matrices $\mathbf{W}_{0}=(\delta _{ij}~\mathbf{0}%
)^{\mathrm{T}}$ and $\widetilde{\mathbf{W}}_{0}=(\mathbf{0\ }\delta _{ij})^{%
\mathrm{T}}$. Reasoning similar to that in \S II.C leads us to the
conclusion that the effective speeds $c_{\alpha }(\kappa _{1})=\lim_{\omega
,k_{1}\rightarrow 0}\omega /k_{1}$ $(\alpha =1,2,3)$ of the three waves with
$\mathbf{k}\equiv k\pmb{\kappa }$ parallel to $\mathbf{a}_{1}$ are the
eigenvalues of the 3$\times $3 matrix
\begin{equation}
\left\langle \left\langle \widetilde{\mathbf{W}}_{0}\cdot (\mathcal{M}_{0}-%
\mathcal{I})^{-1}\mathbf{W}_{0}\right\rangle _{2}\right\rangle _{3}\ \big(%
\mathrm{with}\ \langle \cdot \rangle _{i}\equiv (\ref{6.1})\big).
\label{3D4}
\end{equation}

\section{Numerical implementation\label{N}}

There are several ways to use the above analytical results for calculating
the effective speed. One approach is to transform to Fourier space with
respect to coordinate(s) other than the coordinate of integration in the
monodromy matrix. Consider the 2D case and apply the Fourier expansion $%
f(x_{1},x_{2})=\sum_{n\in \mathbb{Z}}\widehat{f}_{n}(x_{1})e^{2\pi inx_{2}}$
in $x_{2}$ for the functions $f=\mu $ and $\mu ^{-1}$. Then the operator of
multiplying by the function $\mu ^{-1}(x_{1},\cdot )$ and the differential
operator $\mathcal{A}(x_{1})\mathcal{=-\partial }_{2}(\mu (x_{1},\cdot )%
\mathcal{\partial }_{2})$ become matrices%
\begin{align}
& \mu ^{-1}\longmapsto \pmb{\mu }^{-1}(x_{1})=(\widehat{\mu ^{-1}}%
_{n-m})=(\widehat{\mu }_{n-m})^{-1},  \notag \\
& \mathcal{A}\longmapsto \mathbf{A}(x_{1})=4\pi ^{2}(nm\widehat{\mu }%
_{n-m}),\ n,m\in \mathbb{Z},  \label{n1}
\end{align}%
and Eq.\ (\ref{5}) reduces to following form
\begin{align}
c^{2}(\kappa _{1})& =\langle \rho \rangle ^{-1}\widetilde{\mathbf{w}}_{%
\widehat{0}}\cdot (\mathbf{M}_{0}-\mathbf{I})^{-1}\mathbf{w}_{\widehat{0}}\
\mathrm{with\ }  \notag \\
\mathbf{M}_{0}& =%
\begin{matrix}
\widehat{\int }_{0}^{1}%
\end{matrix}%
(\mathbf{I}+\mathbf{Q}_{0}dx_{1}),\ \ \mathbf{Q}_{0}(x_{1})=%
\begin{pmatrix}
\mathbf{0} & \pmb{\mu }^{-1} \\
\mathbf{A} & \mathbf{0}%
\end{pmatrix}%
,\   \label{n3} \\
\widetilde{\mathbf{w}}_{\widehat{0}}& =(\mathbf{0\ }\delta _{0n})^{\mathrm{T}%
},\ \ \mathbf{w}_{\widehat{0}}=(\delta _{0n}\ \mathbf{0})^{\mathrm{T}},
\notag
\end{align}%
where $c(\kappa _{1})=c=$const.\ for any $\pmb{\kappa }$ in the isotropic
case. The above vectors and matrices are, strictly speaking, of infinite
dimension, which needs to be truncated for numerical purposes. In this sense
there is no loss of generality in assuming a smooth $\mu (\mathbf{x})$ in
the course of derivations in \S \ref{2D}. Implementation of Eq.\ (\ref{n3})$%
_{1}$ consists of two steps.

\begin{figure}[h]
\centering
 \includegraphics{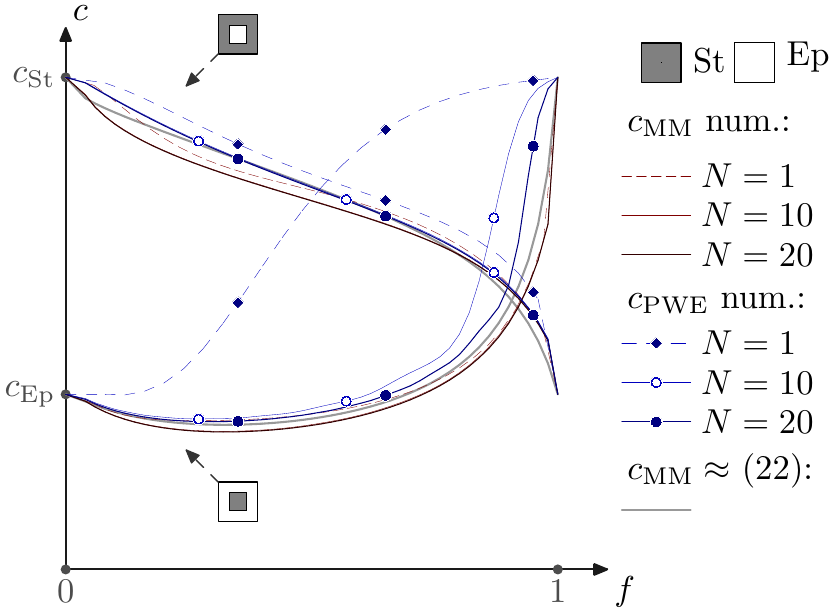}
\caption{Effective speed $c$ versus concentration $f$ of square rods
for 2D St/Ep and Ep/St lattices (see details in the text).}
\label{fig1}
\end{figure}

\noindent \textbf{Step 1.} Calculate the multiplicative integral (\ref{n3})$%
_2$ defining $\mathbf{M}_{0}$. For an arbitrary $\mu ( \mathbf{x}) $, one
way is to use a discretization scheme. Divide the segment $x_{1}\in \lbrack
0,1]$ into $N_{1}$ intervals $[x_{1}^{( i) },x_{1}^{( i+1) })\equiv \Delta
_{i}$, $i=1..N_{1}$, of small enough length. Calculate $2N+1$ Fourier
coefficients $\widehat{\mu }_{n}(x_{1}^{( i) })$, $n=-N..N$ and the $%
(2N+1)\times (2N+1)$ matrices $\mathbf{Q}_{0}(x_{1}^{( i) })$ for each $%
i=1..N_{1},$ and then use the approximate formula $\mathbf{M}%
_{0}=\prod_{i=N_{1}}^{1}\exp \left[ \Delta _{i}\mathbf{Q}_{0}(x_{1}^{( i) })%
\right] $. Recall that $\widehat{\int }$ satisfies the chain rule and is
exactly equal to $\exp ( \Delta _{i}\mathbf{Q}_{0}) $ for $x_{1}\in \Delta
_{i}$ if $\mu ( \mathbf{x}) $ does not depend on $x_{1}$ within $\Delta _{i}$%
. Therefore the calculation is much simpler in the common case of a
piecewise homogeneous unit cell with only a few inclusions of simple shape
(see the example below).

\noindent \textbf{Step 2.} Solve the system $(\mathbf{M}_{0}-\mathbf{I})%
\mathbf{w}_{\widehat{1}}=i\mathbf{w}_{\widehat{0}}$ for unknown $\mathbf{w}_{%
\widehat{1}}$. First remove one zero row and one zero column in the matrix $%
\mathbf{M}_{0}-\mathbf{I}$ (see the remark below (\ref{3})). Then the vector
$\mathbf{w}_{\widehat{1}}$ is uniquely defined and may be found by any
standard method. Note that only a single component of $\mathbf{h}$ is needed
to evaluate $\widetilde{\mathbf{w}}_{\widehat{0}}\cdot \mathbf{w}_{\widehat{1%
}}\mathbf{.}$ Finally dividing by $\langle \rho \rangle $ yields the desired
result (\ref{n3})$_{1}$.

As an example, we calculate the effective shear-wave speed $c$ versus the
volume fraction $f$ of square rods periodically embedded in a matrix
material forming a 2D square lattice with translations parallel to the
inclusion edges. A high-contrast pair of materials is chosen such as steel ($%
\equiv $ St, with $\rho =7.8\ 10^{3}\ $kg/m$^{3}$,\ $\mu =80\ $GPa) and
epoxy ($\equiv $ Ep, with $\rho =1.14\ 10^{3}\ $kg/m$^{3}$,\ $\mu =1.48\ $%
GPa). We consider two conjugated St/Ep and Ep/St configurations,
where the matrix and rod materials are either St and Ep or Ep and
St, respectively. The results are displayed in Fig.\ 1. The curves
$c_{\mathrm{MM}}(f)$ are computed by the present monodromy-matrix
(MM) method, Eq.\ (\ref{n3})$_{1}$,  they are complemented by the
approximation (\ref{M18}). Also shown for comparison are the curves
$c_{\mathrm{PWE}}(f)$ computed from the truncated formula\cite{KAG}
of the conventional PWE method based on a 2D Fourier transform of
(\ref{6}). Calculations are performed for a different
fixed number $2N+1\equiv d$ of the 1D Fourier coefficients of $\mu (\mathbf{x%
})$, which implies $2d\times 2d$ monodromy matrix in\
(\ref{n3})$_{1}$ and, by contrast, $d^{2}\times d^{2}$ matrix in the
PWE formula\cite{KAG}. Apart from this advantage of the MM
calculation, it is also seen to be remarkably more stable - with a
reasonable fit provided already at $N=1$. The difference between the
MM and PWE numerical curves is especially notable for the case of
densely packed steel rods. Interestingly, the MM computation and
estimate both predict a steep fall for $c\left( f\right) $ when a
small concentration $1-f$ of epoxy forms a 'quasi-insulating
network'.  The PWE fails to capture this important physical feature
for reasons described next.

The far superior stability and accuracy of the MM method observed in
Fig.\ 1 can be  explained as follows. The PWE formula\cite{KAG}
implies calculating $M_{11}\approx \sum_{\left\vert
\mathbf{g}\right\vert <d}B_{\mathbf{g}}\left\vert
\mathbf{g}\right\vert ^{-2}\left( \left\vert g_{2}\right\vert
+1\right) ^{-2}+O\left( d^{-1}\right) $ with bounded coefficients
$B_{\mathbf{g}},$ where $\mathbf{g}$ are the 2D
reciprocal lattice  vectors  (we use here that the components of the vector $\widehat{%
\partial _{1}\mu }$ for piecewise constant $\mu (\mathbf{x})$ are of
order $\left( \left\vert g_{2}\right\vert +1\right) ^{-1},$ and that the
matrix corresponding to ${\mathcal{C}}_{0}^{-1}$ is close to
diagonally-dominant and hence its eigenvalues are of   order $\left\vert
\mathbf{g}\right\vert ^{-2}$). Thus the accuracy of the PWE method is
expected to be of order $d^{-1}.$ In contrast, the accuracy of the MM
method, where the 1D Fourier expansion is performed inside a
multiplicative integral that is 'close' to exponential, is expected to be on
the order $e^{-d}.$ This can be understood from the MM equation (\ref{n3})$%
_{1}$ where the $2d\times 2d$ matrix
$(\mathbf{M}_{0}-\mathbf{I})^{-1}$ can be replaced by
$2(\mathbf{M}_{0}-\mathbf{M}_{0}^{-1})^{-1}$ with eigenvalues of  order $e^{-n},~n=1..d$.

\smallskip
\noindent \textbf{Acknowledgement.} This work has been supported by
the grant ANR-08-BLAN-0101-01 and the project SAMM. A.N.N.
acknowledges support from the CNRS.
\renewcommand{\bibsection}{\ \ \ \ \ \ \ \-------------}

\end{document}